\documentclass[aps,prd,twocolumn,amsmath,amssymb,showpacs,nofootinbib]{revtex4}

\usepackage{graphicx}
\usepackage{array,color}
\usepackage{mathrsfs}
\usepackage{longtable,lscape}
\usepackage{txfonts}
\usepackage{amssymb}
\usepackage{indentfirst}
\usepackage{color}
\usepackage{amssymb}
\usepackage{epsfig}
\usepackage{multirow}

\begin{document}
\preprint{APS/123-QED}
\title{Newly observed $\Omega(2012)$ state and strong decays of the low-lying $\Omega$ excitations }
\author{Zuo-Yun Wang$^{1}$}
\author{Long-Cheng Gui$^{1,2,3}$}
\email{guilongcheng@hunnu.edu.cn}
\author{Qi-Fang L{\"{u}}$^{1,2,3}$}
\email{lvqifang@hunnu.edu.cn}
\author{Li-Ye Xiao$^{4,5}$}
\author{Xian-Hui Zhong$^{1,2,3}$}
\email{zhongxh@hunnu.edu.cn}

\affiliation{1) Department of Physics, Hunan Normal University,}
\affiliation{2) Key Laboratory of Low-Dimensional Quantum Structures and Quantum
Control of Ministry of Education, Changsha 410081, China }
\affiliation{3) Synergetic Innovation Center for Quantum Effects and Applications(SICQEA),
Hunan Normal University, Changsha 410081,China}
\affiliation{4) Center of High Energy Physics, Peking University, Beijing 100871,
China}
\affiliation{5) School of Physics and State Key Laboratory of Nuclear Physics and Technology, Peking University, Beijing 100871, China}
\begin{abstract}
Stimulated by the newly discovered $\Omega(2012)$ resonance at Belle II, in this work we
have studied the OZI allowed strong decays of the low-lying $1P$- and $1D$-wave
$\Omega$ baryons within the $^{3}P_{0}$ model. It is found that $\Omega(2012)$
is most likely to be a $1P$-wave $\Omega$ state with $J^P=3/2^{-}$. We also find that
the $\Omega(2250)$ state could be assigned as a $1D$-wave state with $J^{P}=5/2^{+}$.
The other missing $1P$- and $1D$-wave $\Omega$ baryons may have large potentials to
be observed in their main decay channels.
\end{abstract}

\maketitle

\section{\label{sec:level1}INTRODUCTION}

The study of hadron spectrum is an important way for us to understand
strong interactions. For the baryon spectra, the classification based
on $SU(3)_{f}$ flavor symmetry has been achieved a great success. The
$\Omega$ hyperon as a member of baryon decuplet in the quark model
was unambiguously discovered in both production and decay
at BNL about one half century ago~\cite{Barnes:1964pd}.
More excited $\Omega$ baryons should exist as well
according to $SU(6)\times O(3)$ symmetry. In theory,
the mass spectrum of $\Omega$ hyperon has been predicted within many models,
such as the Skyrme model~\cite{Oh2007a}, various constituent quark models \citep{CapstickSimonandIsgur1986,Faustov2015,LoringUlrichandMetschBernardC.andPetry2001,
LiuJianglaiandMcKeownRobertD.andRamsey-Musolf2007,ChaoKuang-TaandIsgurNathanandKarl1981a,
ChenYanandMa2009a,AnC.S.andMetschB.Ch.andZou2013,AnC.S.andZou2014,HayneCameronandIsgur1982,
PervinMuslemaandRoberts2008}, the lattice gauge theory~\citep{EngelGeorgP.andLangC.B.andMohlerDanielandSchafer2013,Liang2015}, and so on.
However, in experiments there are only a few information of the excited $\Omega$ baryons.
In the review of particle physics from the Particle Data Group (PDG),
except for the ground state $\Omega(1672)$, only three possible excited $\Omega$ baryons are listed:
$\Omega(2250)$, $\Omega(2380)$, and $\Omega(2470)$~\cite{Patrignani2016}. Their nature is still
rather uncertain with three- or two-star ratings.
Fortunately, the Belle II experiments offer a great opportunity for our study of the $\Omega$ spectrum.

Very recently, a candidate of excited $\Omega$ baryon, $\Omega(2012)$, was observed by
the Belle II collaboration~\citep{Yelton2018}. The measured mass and width are
\begin{eqnarray}
M&=&2012.4\pm0.7(\textrm{stat})\pm0.6(\textrm{syst})\textrm{ MeV} \nonumber,\\
\Gamma &=& 6.4_{-2.0}^{+2.5}(\textrm{stat})\pm1.6(\textrm{syst})\textrm{ MeV}, \nonumber
\end{eqnarray}
respectively. In various quark models~\citep{CapstickSimonandIsgur1986,Faustov2015,LoringUlrichandMetschBernardC.andPetry2001,
LiuJianglaiandMcKeownRobertD.andRamsey-Musolf2007,ChaoKuang-TaandIsgurNathanandKarl1981a,
ChenYanandMa2009a,AnC.S.andMetschB.Ch.andZou2013,AnC.S.andZou2014,HayneCameronandIsgur1982,
PervinMuslemaandRoberts2008}, the masses of the first orbital ($1P$) excitations of $\Omega$
states are predicted to be $\sim 2.0$ GeV. The newly
observed state $\Omega(2012)$ may be a good candidate of the $1P$-wave $\Omega$ state.
Recently, to study the possible interpretation of $\Omega(2012)$,
its strong decays were calculated with the chiral quark model~\citep{Xiao2018a},
where it was shown that $\Omega(2012)$ could be
assigned to the spin-parity $J^{P}=3/2^{-}$ $1P$-wave state. However,
the spin-parity $J^{P}=1/2^{-}$ state can't be completely ruled out.
Furthermore, the mass and strong decay patten of $\Omega(2012)$ also were studied by QCD sum
rule~\citep{AlievT.M.andAziziK.andSaracY.andSundu2018,Aliev2018}.
As its mass is very close to $\Xi(1530)K$ threshold, there is also
some work considered it as a $J^{P}=3/2^{-}$ $K\Xi^{*}$ molecular
state \citep{Valderrama2018,R.Pavao2018,Lin2018}, or a dynamically generated
state \citep{Huang2012}. In Ref.~\citep{Polyakov2012}, by a flavor $SU(3)$ analysis
the authors suggested $\Omega(2012)$ to be a number of $3/2^{-}$ decuplet baryon if the sum of branching
ratios for the decay $\Omega(2012)\to\Xi\bar{K}\pi$, $\Omega^{-}\pi\pi$
is not too large ($\leqslant70\%$).

In present work, to further reveal the nature
of $\Omega(2012)$ and better understand the properties of the excited $\Omega$ states,
we study the Okubo-Zweig-Iizuka(OZI) allowed two-body strong decays of
the $1P$- and $1D$-wave baryons with the widely used $^{3}P_{0}$ model. The quark model
classification for the $1P$- and $1D$-wave $\Omega$ baryons and their theoretical masses
are listed in Table~\ref{tab:table-1}. The spatial wave functions for the $\Omega$ baryons
are described by harmonic oscillators.
According to our calculations, we find that (i) the newly observed $\Omega(2012)$ resonance
is most likely to be the $1P$-wave $\Omega$ state with spin-parity $J^P=3/2^-$
and the experimental data can be reasonably described. The other $1P$-wave state with $J^P=1/2^-$
might be broader state with a width of dozens of MeV. The $^{3}P_{0}$ results
are consistent with the recent predictions of the chiral quark model~\cite{Xiao2018a}. (ii)
The $\Omega(2250)$ resonance listed in PDG may be a good
candidate of the $J^P=5/2^+$ $1D$ wave state $|56,^410,2,2,5/2^+\rangle$ or $|70,^{2}10,2,2,5/2^{+}\rangle $.
Although the widths of the $D$-wave states predicted within the $^{3}P_{0}$ model
are systematically larger that those predicted with the chiral quark model
~\cite{Xiao2018a}, these states may be observed in their main decay channels
in future experiments for their relatively narrow width.

This paper is organized as follows. Firstly, We give a brief review
of the $^{3}P_{0}$ model in Sec. II. Secondly, we present the numerical
results of strong decay of $1P$- and $1D$-wave $\Omega$ baryon in Sec. III.
Finally, a summary of our results is given in Sec. IV.

\begin{table*}[t]
\caption{\label{tab:table-1}The theory predicted masses (MeV) and spin-flavor-space
wave-functions of $\Omega$ baryons under $SU(6)$ quark model classification
are listed below. We denote the baryon states as $|\textrm{N}_{6},^{2S+1}\textrm{N}_{3},N,L,J^{P}\rangle$
where $N_{6}$ stands for the irreducible representation of spin-flavor
$SU(6)$ group, $N_{3}$ stands for the irreducible representation of
flavor $SU(3)$ group and N, L, $J^{P}$ as principal quantum number,
total orbital angular momentum and spin-parity, respectively\cite{Xiao2018a}.
The $\phi$,$\chi$,$\psi$ denote flavor, spin and spatial wave function, respectively.
The Clebsch-Gorden coefficients of spin-orbital coupling have been omitted.}
\begin{ruledtabular}
\begin{tabular}{cccccccc}
 &  &  & \multicolumn{4}{c}{Theory} & \tabularnewline
\hline
 & States & Wave function & \citep{Faustov2015} & \citep{CapstickSimonandIsgur1986} & \citep{Oh2007a} & \citep{ChaoKuang-TaandIsgurNathanandKarl1981a} & \citep{EngelGeorgP.andLangC.B.andMohlerDanielandSchafer2013}\tabularnewline
\hline
1S & $\left|56,^{4}10,0,0,3/2^{+}\right\rangle $ & $\phi^{s}\chi^{s}\psi_{000}^{s}$ & $1678$ & $1635$ & $1694$ & $1675$ & $1642(17)$\tabularnewline
\hline
1P & $\left|70,^{2}10,1,1,1/2^{-}\right\rangle $ & $\frac{1}{\sqrt{2}}\left(\phi^{s}\chi^{\rho}\psi_{11L_{z}}^{\rho}+\phi^{s}\chi^{\lambda}\psi_{11L_{z}}^{\lambda}\right)$ & $1941$ & $1950$ & $1837$ & $2020$ & $1944(56)$\tabularnewline
 & $\left|70,^{2}10,1,1,3/2^{-}\right\rangle $ & $\frac{1}{\sqrt{2}}\left(\phi^{s}\chi^{\rho}\psi_{11L_{z}}^{\rho}+\phi^{s}\chi^{\lambda}\psi_{11L_{z}}^{\lambda}\right)$ & $2038$ & $2000$ & $1978$ & $2020$ & $2049(32)$\tabularnewline
\hline
1D & $\left|56,^{4}10,2,2,1/2^{+}\right\rangle $ & $\phi^{s}\chi^{s}\psi_{22L_{z}}^{s}$ & $2301$ & $2255$ & $2140$ & $2210$ & $2350(63)/2481(51)$\tabularnewline
 & $\left|56,^{4}10,2,2,3/2^{+}\right\rangle $ & $\phi^{s}\chi^{s}\psi_{22L_{z}}^{s}$ & $2173/2304$ & $2280$ & $2282$ & $2215$ & \multirow{2}{*}{$2470(49)$}\tabularnewline
 & $\left|70,^{2}10,2,2,3/2^{+}\right\rangle $ & $\frac{1}{\sqrt{2}}\left(\phi^{s}\chi^{\rho}\psi_{22L_{z}}^{\rho}+\phi^{s}\chi^{\lambda}\psi_{22L_{z}}^{\lambda}\right)$ & $2173/2304$ & $2345$ & $2282$ & $2265$ & \tabularnewline
 & $\left|56,^{4}10,2,2,5/2^{+}\right\rangle $ & $\phi^{s}\chi^{s}\psi_{22L_{z}}^{s}$ & $2401$ & $2280$ & $\cdots$ & $2225$ & \tabularnewline
 & $\left|70,^{2}10,2,2,5/2^{+}\right\rangle $ & $\frac{1}{\sqrt{2}}\left(\phi^{s}\chi^{\rho}\psi_{22L_{z}}^{\rho}+\phi^{s}\chi^{\lambda}\psi_{22L_{z}}^{\lambda}\right)$ & $2401$ & $2345$ & $\cdots$ & $2265$ & \tabularnewline
 & $\left|56,^{4}10,2,2,7/2^{+}\right\rangle $ & $\phi^{s}\chi^{s}\psi_{22L_{z}}^{s}$ & $2332$ & $2295$ & $\cdots$ & $2210$ & \tabularnewline
\end{tabular}
\end{ruledtabular}
\end{table*}

\section{THE $^{3}P_{0}$ MODEL }

The $^{3}P_{0}$ model, is also called the quark pair creation (QPC)
model. It was first proposed by Micu \citep{Micu1969}, Carlitz and
Kislinger \citep{CarlitzRobertD.andKislinger1970}, and  developed
by the Orsay group and Yaouanc et al \citep{LeYaouancA.andOliverL.andPeneO.andRaynal1974,LeYaouancA.andOliverL.andPeneO.andRaynal1975,LeYaouancA.andOliverL.andPeneO.andRaynal1977a,LeYaouancA.andOliverL.andPeneO.andRaynal1977,LeYaouancA.andOliverL.andPeneO.andRaynal1988,LeYaouanc1973}.
In the model, it assumes that a pair of quarks $q\bar{q}$ is created
from the vacuum with $J^{PC}=0^{++}(^{2S+1}L_{J}={}^{3}P_{0})$ when
the initial hadron A decays, and the quarks from the hadron A regroups
with the created quarks form two daughter hadrons B and C. For baryon
decays, two quarks of the initial baryon regroups with the created
quark to form a daughter baryon, and the remaining one quark regroup
with the created antiquark to form a meson. The process of baryon
decays is shown in Fig. 1.

The transition operator under the $^{3}P_{0}$ model is written as

\begin{eqnarray}
T & = & -3\gamma\underset{m}{\sum}\langle1m;1-m|00\rangle\int d^{3}\vec{p}_{4}d^{3}\vec{p}_{5}\delta^{3}(\vec{p}_{4}+\vec{p}_{5})\nonumber \\
 &  & \times\mathcal{Y}_{1}^{m}(\frac{\vec{p}_{4}-\vec{p}_{5}}{2})\chi_{1,-m}^{45}\varphi_{0}^{45}\omega_{0}^{45}b_{4i}^{\dagger}(\vec{p}_{4})d_{5j}^{\dagger}(\vec{p}_{5}),
\end{eqnarray}

\noindent where the pair-strength $\gamma$ is a dimensionless parameter, $i$ and $j$ are the color
indices of the created quark $\ensuremath{q_{4}}$ and antiquark $\ensuremath{\bar{q}_{5}}$.
$\ensuremath{\varphi_{0}^{45}=(u\bar{u}+d\bar{d}+s\bar{s})/\sqrt{3}}$
and $\ensuremath{\omega_{0}^{45}=\delta_{ij}}$ stand for the flavor
and color singlet, respectively. $\ensuremath{\chi_{1,-m}^{45}}$
is the spin triplet state and $\ensuremath{\mathcal{Y}_{1}^{m}(\vec{p})\equiv\left|\vec{p}\right|Y_{1}^{m}(\theta_{p},\phi_{p})}$
is a solid harmonic polynomial corresponding to the $P$-wave quark
pair. $b_{4i}^{\dagger}(\vec{p}_{4})d_{5j}^{\dagger}(\vec{p}_{5})$
denotes the creation operator in the vacuum.

According to the definition of the mock state \citep{HayneCameronandIsgur1982},
the baryons A and B, and meson C are defined as follows :
\begin{eqnarray}
 &  & \left|A(n_{A}^{2S_{A}+1}L_{A(J_{A},M_{J_{A}})})(\vec{P}_{A})\right\rangle \nonumber \\
 & = & \sqrt{2E_{A}}\underset{M_{L_{A}},M_{S_{A}}}{\sum}\langle L_{A}M_{L_{A}};S_{A}M_{S_{A}}|J_{A}M_{J_{A}}\rangle\nonumber \\
 &  & \times\int\delta^{3}(\vec{p}_{1}+\vec{p}_{2}+\vec{p}_{3}-\vec{P}_{A})d^{3}\vec{p}_{1}d^{3}\vec{p}_{2}d^{3}\vec{p}_{3}\nonumber \\
 &  & \times\Psi_{n_{A}L_{A}M_{L_{A}}}(\vec{p}_{1},\vec{p}_{2},\vec{p}_{3})\chi_{S_{A}M_{S_{A}}}^{123}\varphi_{A}^{123}\omega_{A}^{123}\nonumber \\
 &  & \times\left|q_{1}(\vec{p}_{1})q_{2}(\vec{p}_{2})q_{3}(\vec{p}_{3})\right\rangle,
\end{eqnarray}

\begin{eqnarray}
 &  & \left|B(n_{B}^{2S_{B}+1}L_{B(J_{B}M_{J_{B}})})(\vec{P}_{B})\right\rangle  \nonumber \\
 & = & \sqrt{2E_{B}}\sum_{M_{L_{B}}M_{S_{B}}}\langle L_{B}M_{L_{B}};S_{B}M_{S_{B}}|J_{B}M_{J_{B}}\rangle \nonumber \\
 &  & \times \int\delta^{3}(\vec{p}_{1}+\vec{p}_{2}+\vec{p}_{4}-\vec{P}_{B})d^{3}\vec{p}_{1}d^{3}\vec{p}_{2}d^{3}\vec{p}_{4}\nonumber \\
 &  & \times\Psi_{n_{B}L_{B}M_{L_{B}}}(\vec{p}_{1},\vec{p}_{2},\vec{p}_{4})\chi_{S_{B}M_{S_{B}}}^{124}\varphi_{B}^{124}\omega_{B}^{124}\nonumber \\
 &  & \times \left|q_{1}(\vec{p}_{1})q_{2}(\vec{p}_{2})q_{4}(\vec{p}_{4})\right\rangle ,
\end{eqnarray}

\begin{eqnarray}
 &  & \left|C(n_{C}^{2S_{C}+1}L_{C(J_{C},M_{J_{C}})})(\vec{P}_{C})\right\rangle \nonumber \\
 & = & \sqrt{2E_{C}}\underset{M_{L_{C}},M_{S_{C}}}{\sum}\langle L_{C}M_{L_{C}};S_{C}M_{S_{C}}|J_{C}M_{J_{C}}\rangle\nonumber \\
 &  & \times\int\delta^{3}(\vec{p}_{3}+\vec{p}_{5}-\vec{P}_{C})d^{3}\vec{p}_{3}d^{3}\vec{p}_{5}\nonumber \\
 &  & \times\Psi_{n_{C}L_{C}M_{L_{C}}}(\vec{p}_{3},\vec{p}_{5})\chi_{S_{C}M_{S_{C}}}^{35}\nonumber \\
 &  & \times\varphi_{C}^{35}\omega_{C}^{35}\left|q_{3}(\vec{p}_{3})q_{5}(\vec{p}_{5})\right\rangle,
\end{eqnarray}
where the subscripts 1,2,3 denote the quarks of the initial baryon $A$.
1,2 and 4 denote the quarks of the final baryon $B$,
3 and 5 stand for the quark and antiquark of the final meson $C$, respectively.
$\vec{p}_{i}(i=1,2,3)$ are the momentum of quarks in baryon $A$.
$\vec{p}_{i}(i=1,2,4)$ are the momentum of quarks in baryon $B$.
$\vec{p}_{3}$ and $\vec{p}_{5}$ are the momentum of the quark and
antiquark in meson $C$. $\vec{P}_{A},\vec{P}_B,\vec{P}_C$ denotes the momentum of
state A,B,C. $S_{A},S_{B},S_C$ and $J_{A},J_B,J_C$ represent
the total spin and the total angular momentum of state A,B,C. $\Psi_{n_{A}L_{A}M_{L_{A}}}$
,$\Psi_{n_A L_A M_{L_A}}$and $\Psi_{n_{C}L_{C}M_{L_{C}}}$ denotes the spatial wave functions
of the baryon and meson in momentum space, respectively.

The decay amplitude of the $\Omega$ baryon in $^{3}P_{0}$ is written
as
\begin{widetext}
\begin{eqnarray}
\mathcal{M}^{M_{J_{A}}M_{J_{B}}M_{J_{C}}} & = & \langle B(n_{B}^{2S_{B}+1}L_{B(J_{B},M_{J_{B}})}),C(n_{C}^{2S_{C}+1}L_{C(J_{C},M_{J_{C}})})|T|A(n_{A}^{2S_{A}+1}L_{A(J_{A},M_{J_{A}})}\rangle\nonumber \\
 & = & -3\gamma\sqrt{8E_{A}E_{B}E_{C}}\underset{M_{L_{A}},M_{S_{A}}}{\sum}\underset{M_{L_{B}},M_{S_{B}}}{\sum}\underset{M_{L_{C}},M_{S_{C}},m}{\sum}\langle L_{A}M_{L_{A}};S_{A}M_{S_{A}}|J_{A}M_{J_{A}}\rangle\langle1m;1-m|00\rangle\nonumber \\
 &  & \times\langle L_{B}M_{L_{B}};S_{B}M_{S_{B}}|J_{B}M_{J_{B}}\rangle\langle L_{C}M_{L_{C}};S_{C}M_{S_{C}}|J_{C}M_{J_{C}}\rangle\langle\chi_{S_{B}M_{S_{B}}}^{124}\chi_{S_{C}M_{S_{C}}}^{35}|\chi_{S_{A}M_{S_{A}}}^{123}\chi_{1-m}^{45}\rangle\nonumber \\
 &  & \times\langle\varphi_{B}^{124}\varphi_{C}^{35}|\varphi_{A}^{123}\varphi_{0}^{45}\rangle\times \sum_{\mbox{\tiny $\begin{array}{c}M_{L_\lambda^A},M_{L_\lambda^B}, M_{L_C} \\ M_{L_\rho^A},M_{L_\rho^B} \end{array}$ }} I_{n^{B}_\rho L^B_\rho M_{L_\rho^B}n^{B}_\lambda L^B_\lambda M_{L_\lambda^B}n_{C}L_{C}M_{L_{C}}}^{n^{A}_\rho L_{\rho}^{A},M_{L_{\rho}^{A}},n^A_\lambda L_{\lambda}^{A},M_{L_{\lambda}^{A}},1m}(\vec{P}_{B}),\label{eq:decay amplitudes}
\end{eqnarray}

\begin{eqnarray}
\langle\chi_{S_{B}M_{S_{B}}}^{124}\chi_{S_{C}M_{S_{C}}}^{35}|\chi_{S_{A}M_{S_{A}}}^{123}\chi_{1-m}^{45}\rangle & = & \langle S_{1}M_{S_{1}};S_{2}M_{S_{2}}|S_{12}M_{S_{12}}\rangle\langle S_{12}M_{S_{12}};S_{3}M_{S_{3}}|S_{A}M_{S_{A}}\rangle\nonumber \\
 &  & \times\langle S_{1}M_{S_{1}};S_{2}M_{S_{2}}|S_{12}M_{S_{12}}\rangle\langle\ensuremath{S_{12}}\ensuremath{M_{S_{12}};}\ensuremath{\ensuremath{\ensuremath{\ensuremath{S_{4}M_{S_{4}}|}S_{B}M_{S_{B}}\rangle}}}\nonumber \\
 &  & \times\text{\text{\text{\ensuremath{\langle S_{3}M_{S_{3}};S_{5}M_{S_{5}}|S_{C}M_{S_{C}}\rangle\langle S_{4}M_{S_{4}};S_{5}M_{S_{5}}|10\rangle}}}}.
\end{eqnarray}
\end{widetext}

\noindent In the equation Eq.~(\ref{eq:decay amplitudes}), $\langle\chi_{S_{B}M_{S_{B}}}^{124}\chi_{S_{C}M_{S_{C}}}^{35}|\chi_{S_{A}M_{S_{A}}}^{123}\chi_{1-m}^{45}\rangle$
and $\langle\varphi_{B}^{124}\varphi_{C}^{35}|\varphi_{A}^{123}\varphi_{0}^{45}\rangle$
stand for the spin matrix and the flavor matrix, respectively. $S_{i}$
stand for the spin of $i$th quark and $S_{12}$ stand for the total
spin of 1 and 2 quarks. The pre-factor 3 in front of $\gamma$ arises
from the fact that the three decay processes of the $\Omega$ baryon
in $^{3}P_{0}$ model are equivalent.The summed magnetic quantum numbers are not completely independent of each other.

\begin{figure}
\includegraphics[width=6cm,height=8cm,keepaspectratio]{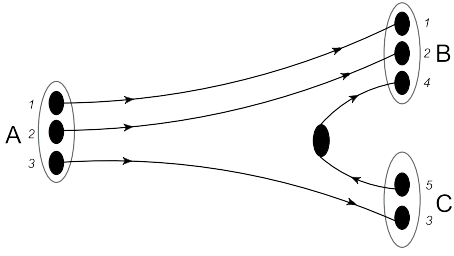}\caption{The decay process of $A\rightarrow B+C$ in the $^{3}P_{0}$ model. \label{fig:decay figure}}

\end{figure}

\noindent The overlap integral in the momentum space is written as
\begin{widetext}
\begin{eqnarray}
I_{n^{B}_\rho L^B_\rho M_{L_\rho^B}n^{B}_\lambda L^B_\lambda M_{L_\lambda^B}n_{C}L_{C}M_{L_{C}}}^{n^{A}_\rho L_{\rho}^{A},M_{L_{\rho}^{A}},n^A_\lambda L_{\lambda}^{A},M_{L_{\lambda}^{A}},1m}(\vec{P}_{B}) & = & \int d^{3}\vec{p}_{1}d^{3}\vec{p}_{2}d^{3}\vec{p}_{3}d^{3}\vec{p}_{4}d^{3}\vec{p}_{5}\delta^{3}(\vec{p}_{1}+\vec{p}_{2}+\vec{p}_{3}-\vec{P}_{A})\delta^{3}(\vec{p}_{4}+\vec{p}_{5})\nonumber \\
 &  & \times\delta^{3}(\vec{p}_{1}+\vec{p}_{2}+\vec{p}_{4}-\vec{P}_{B})\delta^{3}(\vec{p}_{3}+\vec{p}_{5}-\vec{P}_{C})\Psi_{A}(\vec{p}_{1},\vec{p}_{2},\vec{p}_{3})\Psi_{B}^{*}(\vec{p}_{1},\vec{p}_{2},\vec{p}_{4})\Psi_{C}^{*}(\vec{p}_{3},\vec{p}_{5})\nonumber \\
 & = & \int d\vec{p}_{\rho}d\vec{p}_{\lambda}|J|\Psi_{A}(\vec{p}_{\lambda}^{A},\vec{p}_{\rho}^{A})\Psi_{B}(\vec{P}_{B},\vec{p}_{\lambda}^{A},\vec{p}_{\rho}^{A})\Psi_{C}(-\vec{P}_{B},\vec{p}_{\lambda}^{A},\vec{p}_{\rho}^{A})\nonumber \\
 & = & \int d\vec{p}_{\rho}^{A}d\vec{p}_{\lambda}^{A}d\Omega_{\rho}^{A}d\Omega_{\lambda}^{A}|J|\phi_{n_{\rho}^A L_{\rho}^{A}}(\vec{p}_{\rho}^{A})Y_{L_{\rho}^{A}M_{L_{\rho}^{A}}}(\Omega_{\rho}^{A})\phi_{n^{A}_\lambda L_{\lambda}^{A}}(\vec{p}_{\lambda}^{A})Y_{L_{\lambda}^{A}M_{L_{\lambda}^{A}}}(\Omega_{\lambda}^{A})\nonumber \\
 &  & \times\phi_{n^{B}_\rho L_{\rho}^{B}}(\vec{p}_{\rho}^{A})Y_{L_{\rho}^{B}M_{L_{\rho}^{B}}}(\Omega_{\rho}^{B})\phi_{n^{B}_\lambda L_{\lambda}^{B}}(\vec{p}_{\lambda}^{A}-\sqrt{\frac{3}{2}}\frac{2m_{3}}{2m_{3}+m_{q}}\vec{P}_{B})Y_{L_{\lambda}^{B}M_{L_{\lambda}^{B}}}(\Omega_{\lambda}^{B})\nonumber \\
 &  & \times\phi_{n_{C}L_{C}}(\frac{m_{3}}{m_{3}+m_{q}}\vec{P}_{B}-\sqrt{\frac{2}{3}}\vec{p}_{\lambda}^{A})Y_{L_{C}M_{L_{C}}}(\Omega^{C})\nonumber \\
 &  & \times\mathcal{Y}_{1}^{m}(\vec{P}_{B}-\sqrt{\frac{2}{3}}\vec{p}_{\lambda}^{A}),\label{eq:momentum space integral}
\end{eqnarray}
where $\vec{p}_{\rho}=\frac{1}{\sqrt{2}}(\frac{m_{2}\vec{p}_{1}-m_{1}\vec{p}_{2}}{m_{1}+m_{2}})$
and $\vec{p}_{\lambda}=\sqrt{\frac{2}{3}}(\frac{m_{3}(\vec{p}_{1}+\vec{p}_{2})-(m_{1}+m_{2})\vec{p}_{3}}{m_{1}+m_{2}+m_{3}})$
stand for the momentum corresponding to $\rho$ and $\lambda$ Jacobi
coordinates in the center mass frame of baryon A, and $|J|$ stand for
the Jacobi determinant which determined by the definition of $p_{\rho}$,
$p_{\lambda}$. $m_{i}$ denote the mass of $i$th quark and $m_{q}$
denote the mass of the created quark pairs. the $n_{\rho}$ and $L_{\rho}$
denote the nodal and orbital angular momentum between the 1,2 quarks
(see Fig. \ref{fig:decay figure}), while the $n_{\lambda}$ and $L_{\lambda}$
stand for the nodal and orbital angular momentum between the 1,2 quarks
system and the 3 quark (see Fig. \ref{fig:decay figure}).
\end{widetext}

In the calculations, simple harmonic oscillator (SHO) wave functions is employed as the
hadron wave function. The momentum space wave function of baryon is

\begin{eqnarray}
\Psi_{A}(\vec{P}_{A}) & = & N\psi_{n_{\rho_{A}}l_{\rho_{A}}M_{l_{\rho_{A}}}}(\vec{p}_{\rho_{A}})\psi_{n_{\lambda_{A}}l_{\lambda_{A}}M_{l_{\lambda_{A}}}}(\vec{p}_{\lambda_{A}})\nonumber \\
\psi_{nlM_{l}}(\vec{p}) & = & (-1)^{n}(-i)^{l}\left[\frac{2^{l+2}}{\sqrt{\pi}(2l+1)!!}\right]^{\frac{1}{2}}\nonumber \\
 &  & \times(\frac{1}{\beta})^{l+\frac{3}{2}}\textrm{exp}(-\frac{\vec{p}^{2}}{2\beta^{2}})L_{n}^{l+1/2}(\frac{\vec{p}^{2}}{\beta^{2}})\mathcal{Y}_{l}^{m}(\vec{p}),
\end{eqnarray}

where $N$ stand for a normalization coefficient of the wave function
and $L_{n}^{l+1/2}(\frac{\vec{p}^{2}}{\beta^{2}})$ is the Laguerre
polynomial function. The Clebsch-Gorden coefficients of $l_\rho,l_\lambda$ coupling  are equal to 1 in our case.

The ground state wave function of a meson in the momentum
space is
\begin{equation}
\Psi_{0,0}=\left[\frac{R^{2}}{\pi}\right]^{\frac{3}{4}}\textrm{exp}(-\frac{R^{2}\vec{p}_{ab}^{2}}{2}),
\end{equation}
where the $\vec{p}_{ab}$ stands for the relative momentum
between the quark and antiquark in the meson. As all hadrons in the
final states are $S$-wave in this work, Eq.~(\ref{eq:momentum space integral})
can be further expressed as follows

\begin{equation}
\Pi(L_{\rho}^{A},M_{L_{\rho}^{A}},L_{\lambda}^{A},M_{L_{\lambda}^{A}},m)\equiv I_{n^{B}_\rho L^B_\rho M_{L_\rho^B}n^{B}_\lambda L^B_\lambda M_{L_\lambda^B}n_{C}L_{C}M_{L_{C}}}^{n^{A}_\rho L_{\rho}^{A},M_{L_{\rho}^{A}},n^A_\lambda L_{\lambda}^{A},M_{L_{\lambda}^{A}},1m}
\end{equation}
the expressions of $\Pi(L_{\rho_{A},}M_{L_{\rho}^{A}},L_{\lambda_{A},}M_{L_{\lambda}^{A}},m)$
and harmonic oscillator wave function for the $S$-wave, $P$-wave,
$D$-wave $\Omega$ baryons are collected in the appendix.

The decay width $\Gamma$ of the process $A\to B+C$ is
\begin{eqnarray}
\Gamma=\pi^{2}\frac{\left|\vec{p}\right|}{m_{A}^{2}}\frac{1}{2J_{A}+1}\underset{M_{J_{A}},M_{J_{B}},M_{J_{C}}}{\sum}\left|\mathcal{M}^{M_{J_{A}}M_{J_{B}}M_{J_{C}}}\right|^{2},\label{eq:decay width}
\end{eqnarray}
where $J_{A}$ are the total angular momentum of the initial baryon
$A$. $\vec{p}$ is the momentum of the final baryon in the center
of mass frame of the initial baryon $A$
\begin{equation}
\left|\vec{p}\right|=\frac{\sqrt{\left[m_{A}^{2}-(m_{B}-m_{C})^{2}\right]\left[m_{A}^{2}-(m_{B}+m_{C})^{2}\right]}}{2m_{A}},
\end{equation}
where $m_{A}$, $m_{B}$ and $m_{C}$ are the mass of the initial
and final hadrons.

In order to partly remedy the inadequacy of the nonrelativistic wave
function as the relative momentum $\vec{p}$ increases\citep{Li:1995si,Zhao:1998fn,Zhong:2007fx,Zhong:2008kd,Gui:2018rvv},
the decay amplitude is written as
\begin{equation}
\mathcal{M}(\vec{p})\rightarrow\gamma_{f}\mathcal{M}(\gamma_{f}\vec{p}),
\end{equation}
where $\gamma_{f}$ denotes a commonly Lorentz boost factor,
$\gamma_{f}=m_{B}/E_{B}$. In most decays, the three momenta carried
by the final state baryons are relatively small, which means the nonrelativistic
prescription is reasonable and the corrections from the Lorentz boost
is not drastic.

\section{numerical results and analysis}

In our calculations, we adopt $m_{u}=m_{d}=350$ MeV, $m_{s}=450$
MeV for the constituent quark masses. The masses of the $\Omega$
baryons listed in Table \ref{tab:table-1}, the masses of $K$ mesons
and $\Xi$ baryons are taken from the PDG \citep{Patrignani2016}.
The quantum numbers involved in the calculations are listed in Tab
\ref{tab:table-2}. Due to the orthogonal relationship of the wave
functions, only the $\lambda$ excited mode contributes. There are
three harmonic oscillator parameters, the $\beta_{\rho}$ and $\beta_{\lambda}$
and $R$ in baryon and meson wave functions, respectively. We adopt
$R=2.5$ GeV$^{-1}$ for $K$ mesons \citep{GodfreyStephenandMoats2016}.
The parameter $\beta_{\rho}$ of the $\rho$-mode excitation between
the 1,2 quarks (see Fig. \ref{fig:decay figure}) is taken as $\beta_{\rho}=0.4$
GeV \citep{Xiao:2017udy}. The $\beta_{\lambda}$ is obtained with
the relation~\cite{Zhong:2007gp}:
\begin{equation}
\beta_{\lambda}=\left(\frac{3m_{3}}{2m_{1}+m_{3}}\right)^{\frac{1}{4}}\beta_{\rho}.
\end{equation}
For the quark pair creation strength from the vacuum, we take as those
in Ref.~\citep{GodfreyStephenandMoats2016}, $\gamma=6.95$.

\begin{table}
\caption{\label{tab:table-2}The Quantum numbers of 1$P$-wave and 1$D$-wave
$\Omega$ baryons. Due to the orthogonal relationship of the wave functions,
only the $\lambda$ excited mode contributes. }
\begin{ruledtabular}
\begin{tabular}{ccccccccc}
States & $J$ & $n_{\rho}$ & $l_{\rho}$ & $n_{\lambda}$ & $l_{\lambda}$ & $L$ & $S_{\rho}$ & $S$\tabularnewline
$\left|70,^{2}10,1,1,1/2^{-}\right\rangle $ & $\frac{1}{2}$ & $0$ & $0$ & $0$ & $1$ & $1$ & $1$ & $\frac{1}{2}$\tabularnewline
$\left|70,^{2}10,1,1,3/2^{-}\right\rangle $ & $\frac{3}{2}$ & $0$ & $0$ & $0$ & $1$ & $1$ & $1$ & $\frac{1}{2}$\tabularnewline
$\left|56,^{4}10,2,2,1/2^{+}\right\rangle $ & $\frac{1}{2}$ & $0$ & $0$ & $0$ & $2$ & $2$ & $1$ & $\frac{3}{2}$\tabularnewline
$\left|56,^{4}10,2,2,3/2^{+}\right\rangle $ & $\frac{3}{2}$ & $0$ & $0$ & $0$ & $2$ & $2$ & $1$ & $\frac{3}{2}$\tabularnewline
$\left|70,^{2}10,2,2,3/2^{+}\right\rangle $ & $\frac{3}{2}$ & $0$ & $0$ & $0$ & $2$ & $2$ & $1$ & $\frac{1}{2}$\tabularnewline
$\left|56,^{4}10,2,2,5/2^{+}\right\rangle $ & $\frac{5}{2}$ & $0$ & $0$ & $0$ & $2$ & $2$ & $1$ & $\frac{3}{2}$\tabularnewline
$\left|70,^{2}10,2,2,5/2^{+}\right\rangle $ & $\frac{5}{2}$ & $0$ & $0$ & $0$ & $2$ & $2$ & $1$ & $\frac{1}{2}$\tabularnewline
$\left|56,^{4}10,2,2,7/2^{+}\right\rangle $ & $\frac{7}{2}$ & $0$ & $0$ & $0$ & $2$ & $2$ & $1$ & $\frac{3}{2}$\tabularnewline
\end{tabular}
\end{ruledtabular}

\end{table}

\subsection{The $1 P$-wave states }

According to the $SU(6)$ supermultiplet classification (see table \ref{tab:table-1}),
there are two 1$P$-wave states with $J^{P}=\frac{1}{2}^{-}$ and $J^{P}=\frac{3}{2}^{-}$,
i.e. $\left|70,^{2}10,1,1,1/2^{-}\right\rangle $ and
$\left|70,^{2}10,1,1,3/2^{-}\right\rangle $. The mass of the newly observed
$\Omega(2012)$ state is close to the $1P$-wave $\Omega$ baryon predicted
in various quark models (see table \ref{tab:table-1}). Assuming $\Omega(2012)$ as
a candidate of the $1P$-wave $\Omega$ baryons, we calculate the OZI-allowed two body strong decays in the $^{3}P_{0}$ model,
and list our results in table \ref{tab:table-3}.

It is found that if one
assigns $\Omega(2012)$ as the $J^{P}=\frac{1}{2}^{-}$ state
$\left|70,^{2}10,1,1,1/2^{-}\right\rangle $, the width is predicted to be
\begin{equation}
\Gamma_{\textrm{total}}^{\textrm{th}}\simeq 43 \textrm{ MeV},
\end{equation}
which is too large to be comparable with the width of $\Omega(2012)$.
This width predicted in the $^{3}P_{0}$ model is about a factor 3
larger than that predicted within the chiral quark model~\citep{Xiao2018a}.

On the other hand, assigning $\Omega(2012)$ as the $J^{P}=\frac{3}{2}^{-}$ state $\left|70,^{2}10,1,1,3/2^{-}\right\rangle $,
we find that the width
\begin{equation}
\Gamma_{\textrm{total}}^{\textrm{th}}\simeq 8\textrm{ MeV},
\end{equation}
and the branching fraction ratio
\begin{equation}
R^{\mathrm{th}}=\frac{\Gamma\left[\left|70,^{2}10,1,1,3/2^{-}\right\rangle \rightarrow\Xi^{0}K^{-}\right]}{\Gamma\left[\left|70,^{2}10,1,1,3/2^{-}\right\rangle \rightarrow\Xi^{-}\bar{K}{}^{0}\right]}\simeq1.1,
\end{equation}
are consistent with the measured width $\Gamma^{\mathrm{exp}} = 6.4_{-2.0}^{+2.5}(\textrm{stat})\pm 1.6(\textrm{syst})$ and ratio $R^{\mathrm{exp}}=1.2\pm 0.3$ for $\Omega(2012)$. These $^{3}P_{0}$ model predictions are compatible with
those predicted within the chiral quark model~\citep{Xiao2018a}.

\begin{table}
\caption{\label{tab:table-3}The total decay widths (MeV) of the
$\left|70,^{2}10,1,1,1/2^{-}\right\rangle $ and $\left|70,^{2}10,1,1,3/2^{-}\right\rangle $
states with mass $M=2012\textrm{ MeV}$. $\Gamma_{\textrm{total}}^{\textrm{th}}$
denotes the total decay width and $\mathfrak{B}$ stands for the radio
of the branching fraction $\Gamma[\Xi^{0}K^{-}]/\Gamma[\Xi^{-}\bar{K}^{0}]$.
The results of Ref. \cite{Xiao2018a} are also listed for a comparison.
The units of widths is MeV.}
\begin{ruledtabular}
\begin{tabular}{ccccc}
    States                                      & $\Gamma_{\textrm{total}}^{\textrm{th}}$ & $\Gamma_{\textrm{total}}^{\textrm{th}}$ \cite{Xiao2018a} & $\mathfrak{B}$  & $\mathfrak{B}$\cite{Xiao2018a} \tabularnewline
$\left|70,^{2}10,1,1,1/2^{-}\right\rangle $     & $43.0$                                  & $15.2$                                                   & $0.96$          & $0.95$ \tabularnewline
$\left|70,^{2}10,1,1,3/2^{-}\right\rangle $     & $8.19$                                  & $6.64$                                                   & $1.11$          & $1.12$ \tabularnewline
\end{tabular}
\end{ruledtabular}
\end{table}

\subsection{The 1 $D$-wave states }

According to the quark model classification, there are six
1$D$-wave states. Their masses are predicted to be in the range of
2.2-2.3 GeV in various models (see {tab:table-1}).
Their OZI allowed two-body strong decay channels are $\Xi K$
and $\Xi(1530)K$. With the masses predicted in Ref.~\citep{ChaoKuang-TaandIsgurNathanandKarl1981a},
we study the strong decay processes of 1$D$-wave states into both $\Xi K$
and $\Xi(1530)K$ channels, and collect their partial decay widths
in Table~\ref{tab:table-4}.

It is interesting to find that the two $J^{P}=5/2^{+}$
states $ |56,^{4}10,2,2,5/2^{+} \rangle $ and $ |70,^{2}10,2,2,5/2^{+} \rangle $
may be good candidates of $\Omega(2250)$ listed in the review book of PDG~\cite{Patrignani2016}.
(i) The mass of $\Omega(2250)$ is close the predictions of $\left|56,^{4}10,2,2,5/2^{+}\right\rangle $
and $ |70,^{2}10,2,2,5/2^{+} \rangle $ in the quark model~\citep{ChaoKuang-TaandIsgurNathanandKarl1981a}.
(ii) The measured width of $\Omega(2250)$, $\Gamma=55\pm18$ MeV, is close to the theoretical predictions, $\Gamma\sim 80/50$ MeV for
$ |56,^{4}10,2,2,5/2^{+} \rangle $ and $ |70,^{2}10,2,2,5/2^{+} \rangle $, respectively. (iii)
The decays modes of $ |56,^{4}10,2,2,5/2^{+} \rangle $ and $ |70,^{2}10,2,2,5/2^{+} \rangle $
are dominated by $\Xi(1530)K$, which is also consistent with the fact that the $\Omega(2250)$
was seen in the $\Xi(1530)K$ and $\Xi^{-}\pi^{+}K^{-}$ channels. In~\citep{Xiao2018a},
we predicted that $\Omega(2250)$ is more likely to be the $J^{P}=5/2^+$
$ |56,^{4}10,2,2,5/2^{+} \rangle $ with the chiral quark model, and the width of $ |70,^{2}10,2,2,5/2^{+} \rangle $ is predicted to
be $\sim 12$ MeV, which is about a factor 4 smaller than the width of $\Omega(2250)$.

The $J^P=3/2^+$ state $|70,^{2}10,2,2,3/2^{+}\rangle $ is a narrow state with a width of
$\Gamma\simeq 23$ MeV, and has comparable decay rates into $\Xi K$
and $\Xi(1530)K$ channels. These predictions of the $^3P_0$ model
are consistent with those in the chiral quark model~\citep{Xiao2018a}.
While the other $J^P=3/2^+$ state $|56,^{4}10,2,2,3/2^{+}\rangle $ is found
to be a broad state with a width of $\Gamma\simeq 130$ MeV, the partial width ratio between
$\Xi K$ and $\Xi(1530)K$ is predicted to be
\begin{equation}
\frac{\Gamma [\Xi K]}{\Gamma [\Xi(1530)K]}\simeq1.4.
\end{equation}
The predicted width for $|56,^{4}10,2,2,3/2^{+}\rangle $ in this work is about a factor 3
larger than that predicted with the chiral quark model~\citep{Xiao2018a}.

The $J^P=1/2^+$ state $|56,^{4}10,2,2,1/2^{+}\rangle $ is the broadest state in the
1$D$-wave states. It has a width of $\Gamma\sim170$ MeV, and mainly decays into
the $\Xi K$ channel. However, in the chiral quark model a relatively
narrower width $\Gamma\sim 56$ MeV is given for $|56,^{4}10,2,2,1/2^{+}\rangle $~\citep{Xiao2018a}.

The $J^P=7/2^+$ state $|56,^{4}10,2,2,7/2^{+}\rangle $ may be a narrow state with a width of
$\Gamma\sim 30$ MeV. This state mainly decays into the $\Xi K$ channel. The
decay properties of $|56,^{4}10,2,2,7/2^{+}\rangle $ predicted in this work are consistent
with those of chiral quark model in Ref.~\citep{Xiao2018a}.

As a whole most of the 1$D$-wave states has a relatively narrow width,
they has potentials to be observed in their dominant decay modes.
The $\Omega(2250)$ resonance may be assigned to the $J^{P}=5/2^{+}$
state $ |56,^{4}10,2,2,5/2^{+} \rangle $ or $ |70,^{2}10,2,2,5/2^{+} \rangle $.
Although the decay widths predicted for the $D$-wave states within the chiral quark model and $^3P_0$ model
show some differences, the partial width ratios of $\Gamma[\Xi K]/\Gamma[\Xi(1530)K]$ predicted within these two
models are in a reasonable agreement with each other.

\begin{table*}[t]
\caption{\label{tab:table-4}The partial and total decay widths (MeV) of the
$1D$-wave states. $\Gamma_{\textrm{total}}^{\textrm{th}}$ denotes
the total decay width and $\mathfrak{B}$ stands for the radio of
the branching fraction $\Gamma[\Xi K]/\Gamma[\Xi(1530)K]$.
The results of Ref. \cite{Xiao2018a} are also shown for a comparison .
The widths and masses are in units of MeV.}
\begin{ruledtabular}
\begin{tabular}{cccccccccc}
    States                                      & Mass\citep{ChaoKuang-TaandIsgurNathanandKarl1981a}   & $\Gamma[\Xi K]$ & $\Gamma[\Xi K]$\cite{Xiao2018a}  & $\Gamma[\Xi(1530) K]$ & $\Gamma[\Xi(1530) K]$ \cite{Xiao2018a} & $\Gamma_{total}^{th}$ & $\Gamma_{total}^{th}$\cite{Xiao2018a} & $\mathcal{B}$ & $\mathcal{B}$\cite{Xiao2018a}\\
         $\left|56,^{4}10,2,2,1/2^{+}\right\rangle $ & $2210$ & $154 $          & $51.8$                           & $16.5$                & $4.53$                                 & $171 $                & $56.3$                                & $9.36$        & $11.4$       \\
         $\left|56,^{4}10,2,2,3/2^{+}\right\rangle $ & $2215$ & $76.8$          & $25.8$                           & $55.4$                & $15.7$                                 & $132 $                & $41.5$                                & $1.39$        & $1.64$       \\
         $\left|56,^{4}10,2,2,5/2^{+}\right\rangle $ & $2225$ & $7.78$          & $6.58$                           & $77.2$                & $22.6$                                 & $84.9$                & $29.2$                                & $0.10$        & $0.29$       \\
         $\left|56,^{4}10,2,2,7/2^{+}\right\rangle $ & $2210$ & $31.7$          & $26.2$                           & $2.94$                & $1.51$                                 & $34.7$                & $27.7$                                & $10.8$        & $17.4$      \\
         $\left|70,^{2}10,2,2,3/2^{+}\right\rangle $ & $2265$ & $9.04$          & $7.40$                           & $14.3$                & $11.9$                                 & $23.4$                & $20.9$                                & $0.63$        & $0.62$       \\
         $\left|70,^{2}10,2,2,5/2^{+}\right\rangle $ & $2265$ & $4.34$          & $0.99$                           & $50.0$                & $11.6$                                 & $54.4$                & $13.4$                                & $0.09$        & $0.08$       \\
\end{tabular}
\end{ruledtabular}
\end{table*}

\section{SUMMARY}

Stimulated by the newly discovered $\Omega(2012)$ resonance at Belle II, in this work we
have studied the OZI allowed strong decays of $1P$- and $1D$-wave
$\Omega$ baryons within the $^{3}P_{0}$ model.

It is found that the newly observed state $\Omega(2012)$ favors the $1P$-wave
$\Omega$ state with $J^P=3/2^-$, $|70,^{2}10,1,1,3/2^{-}\rangle $. Both the decay
mode and decay width are consistent with the observations. The other $1P$-wave
$\Omega$ state with $J^P=1/2^-$, $|70,^{2}10,1,1,1/2^{-}\rangle $, might have
a relatively broad width of $\mathcal{O}(10)$ MeV. This $J^P=1/2^-$ state
should be observed in the $\Xi^0K^-$ and $\Xi^-\bar{K}^0$ channels as well.

In the $1D$-wave $\Omega$ states, it is found that the $\Omega(2250)$ state
may favor the $J^{P}=5/2^{+}$ state $ |56,^{4}10,2,2,5/2^{+} \rangle $
or $ |70,^{2}10,2,2,5/2^{+} \rangle $. These two $J^{P}=5/2^{+}$ states
dominantly decay into the $\Xi(1530)K$ channel, and have a similar decay
width to that of $\Omega(2250)$. Due to a large uncertainty of the width of $\Omega(2250)$,
we can't distinguish it belongs to the 70 multiplet or 56 multiplet. Future experiment
information will help to clarify this issue. For the other $1D$-wave $\Omega$
baryons, we recommend looking for the $J^{P}=1/2^{+}$ and $J^{P}=7/2^{+}$
states in the $\Xi K$ decay channel and looking for the $J^{P}=3/2^{+}$ states in both $\Xi K$
and $\Xi(1530)K$ decay channels in future experiments.

\begin{acknowledgments}
This work is supported by the National Natural Science Foundation
of China under Grants No. 11405053, No.11775078, No.U1832173, No.11705056. This work is also
in part supported by China Postdoctoral Science Foundation under Grant
No. 2017M620492.
\end{acknowledgments}

\appendix

\section{THE HARMONIC OSCILLATOR WAVE FUNCTIONS }

For the $S$-wave omega baryon, the harmonic oscillator wave
function is

\begin{eqnarray}
\Psi(0,0,0,0) & = & \left(\frac{4}{\sqrt{\pi}}\right)^{\frac{1}{2}}\left(\frac{1}{\beta_{\rho}}\right)^{\frac{3}{2}}y_{0,0}(\vec{p}_{\rho})e^{-\frac{\vec{p}_{\rho}^{2}}{2\beta_{\rho}^{2}}-\frac{\vec{p}_{\lambda}^{2}}{2\beta_{\lambda}^{2}}}\nonumber \\
 &  & \times\left(\frac{4}{\sqrt{\pi}}\right)^{\frac{1}{2}}\left(\frac{1}{\beta_{\lambda}}\right)^{\frac{3}{2}}y_{0,0}(\vec{p}_{\lambda})
\end{eqnarray}

For the $P$-wave omega baryon, the harmonic oscillator wave
function is

\begin{eqnarray}
\Psi(0,0,1,m_{l\lambda}) & = & -i\left(\frac{4}{\sqrt{\pi}}\right)^{\frac{1}{2}}\left(\frac{1}{\beta_{\rho}}\right)^{\frac{3}{2}}y_{0,0}(\vec{p}_{\rho})e^{-\frac{\vec{p}_{\rho}^{2}}{2\beta_{\rho}^{2}}-\frac{\vec{p}_{\lambda}^{2}}{2\beta_{\lambda}^{2}}}\nonumber \\
 &  & \times\left(\frac{8}{3\sqrt{\pi}}\right)^{\frac{1}{2}}\left(\frac{1}{\beta_{\lambda}}\right)^{\frac{5}{2}}y_{1,m_{l\lambda}}(\vec{p}_{\lambda}),
\end{eqnarray}

\begin{eqnarray}
\Psi(1,m_{l\rho},0,0) & = & -i\left(\frac{8}{3\sqrt{\pi}}\right)^{\frac{1}{2}}\left(\frac{1}{\beta_{\rho}}\right)^{\frac{5}{2}}y_{1,m_{l\rho}}(\vec{p}_{\rho})e^{-\frac{\vec{p}_{\rho}^{2}}{2\beta_{\rho}^{2}}-\frac{\vec{p}_{\lambda}^{2}}{2\beta_{\lambda}^{2}}}\nonumber \\
 &  & \times\left(\frac{4}{\sqrt{\pi}}\right)^{\frac{1}{2}}\left(\frac{1}{\beta_{\lambda}}\right)^{\frac{3}{2}}y_{0,0}(\vec{p}_{\lambda}),
\end{eqnarray}

For the $D$-wave omega baryon, the harmonic oscillator wave
function is

\begin{eqnarray}
\Psi(0,0,2,m_{l\lambda}) & = & -\left(\frac{4}{\sqrt{\pi}}\right)^{\frac{1}{2}}\left(\frac{1}{\beta_{\rho}}\right)^{\frac{3}{2}}y_{0,0}(\vec{p}_{\rho})e^{-\frac{\vec{p}_{\rho}^{2}}{2\beta_{\rho}^{2}}-\frac{\vec{p}_{\lambda}^{2}}{2\beta_{\lambda}^{2}}}\nonumber \\
 &  & \times\left(\frac{16}{15\sqrt{\pi}}\right)^{\frac{1}{2}}\left(\frac{1}{\beta_{\lambda}}\right)^{\frac{7}{2}}y_{2,m_{l\lambda}}(\vec{p}_{\lambda}),
\end{eqnarray}

\begin{eqnarray}
\Psi(2,m_{l\rho},0,0) & = & -\left(\frac{16}{15\sqrt{\pi}}\right)^{\frac{1}{2}}\left(\frac{1}{\beta_{\rho}}\right)^{\frac{7}{2}}y_{2,m_{l\rho}}(\vec{p}_{\rho})e^{-\frac{\vec{p}_{\rho}^{2}}{2\beta_{\rho}^{2}}-\frac{\vec{p}_{\lambda}^{2}}{2\beta_{\lambda}^{2}}}\nonumber \\
 &  & \times(\frac{4}{\sqrt{\pi}})^{\frac{1}{2}}(\frac{1}{\beta_{\lambda}})^{\frac{3}{2}}y_{0,0}(\vec{p}_{\lambda}),
\end{eqnarray}
\linebreak{}

\noindent Where $\vec{p}_{\rho}=\frac{1}{\sqrt{2}}(\vec{p}_{1}-\vec{p}_{2})$
and $\vec{p}_{\lambda}=\sqrt{\frac{1}{6}}(\vec{p}_{1}+\vec{p}_{2}-2\vec{p}_{3})$
are obtained in the relative Jacobi coordinates. The $y_{l,m_{l}}(\vec{p})$
is the solid harmonic polynomial.

The wave function of the meson in our calculation is

\begin{equation}
\Psi(0,0)=\left(\frac{R^{2}}{\pi}\right)^{\frac{3}{4}}e^{-\frac{(\vec{p}_{3}-\vec{p}_{5})^{2}R^{2}}{8}}
\end{equation}

\noindent Here $\vec{p}_{C}=\frac{\vec{p}_{3}-\vec{p}_{5}}{2}$, $R=\frac{1}{\beta_{C}}$.

\section{THE MOMENTUM SPACE INTEGRATION}

The momentum space integration $\Pi(L^A_{\rho },M^A_{\rho},L^A_{\lambda},M^A_{\lambda },m)$
are presented in the following
\begin{equation}
\Pi(0,0,0,0,0)=\beta\left|\vec{p}\right|\Delta_{0,0}.
\end{equation}

For the $P$-wave omega baryon decay,

\begin{equation}
\Pi(0,0,1,0,0)=\left(\frac{1}{\sqrt{6}\lambda_{2}}-\frac{\lambda_{3}}{2\lambda_{2}}\beta\left|\vec{p}\right|^{2}\right)\Delta_{0,1},
\end{equation}

\begin{align}
\Pi(0,0,1,1,-1) & =-\frac{1}{\sqrt{6}\lambda_{2}}\Delta_{0,1}\nonumber \\
 & =\Pi(0,0,1,-1,1).
\end{align}

For the $D$-wave omega baryon decay,

\begin{equation}
\Pi(0,0,2,0,0)=\left(\frac{\lambda_{3}^{2}}{2\lambda_{2}^{2}}\beta\left|\vec{p}\right|^{3}-\frac{\sqrt{6}\lambda_{3}}{3\lambda_{2}^{2}}\left|\vec{p}\right|\right)\Delta_{0,2},
\end{equation}

\begin{align}
\Pi(0,0,2,1,-1) & =\frac{\lambda_{3}}{\sqrt{2}\lambda_{2}^{2}}\left|\vec{p}\right|\Delta_{0,2}\nonumber \\
 & =\Pi(0,0,2,-1,1).
\end{align}

Here,

\begin{equation}
\lambda_{1}=\frac{1}{2\beta_{\rho}^{2}}+\frac{1}{2\beta_{\rho}^{'2}},\text{\quad}\lambda_{2}=\frac{1}{2\beta_{\lambda}^{2}}+\frac{1}{2\beta_{\lambda}^{'2}}+\frac{R^{2}}{3},
\end{equation}

\begin{equation}
\lambda_{3}=\frac{\sqrt{6}m_{3}}{(2m_{3}+m_{5})\beta_{\lambda}^{'2}}+\frac{m_{3}}{m_{3}+m_{5}}\frac{\sqrt{6}R^{2}}{3},
\end{equation}

\begin{equation}
\lambda_{4}=\frac{3m_{3}^{2}}{(2m_{3}+m_{5})^{2}\beta_{\lambda}^{'2}}+\frac{m_{3}^{2}}{(m_{3}+m_{5})^{2}}\frac{R^{2}}{2},
\end{equation}

\begin{equation}
\beta=1-\frac{\lambda_{3}}{\sqrt{6}\lambda_{2}},
\end{equation}

and

\begin{eqnarray}
\Delta_{0,0} & = & \left(\frac{1}{\pi\beta_{\rho}^{'2}}\right)^{\frac{3}{4}}\left(\frac{1}{\pi\beta_{\lambda}^{'2}}\right)^{\frac{3}{4}}\left(\frac{R^{2}}{\pi}\right)^{\frac{3}{4}}\left(\frac{\pi^{2}}{\lambda_{1}\lambda_{2}}\right)^{\frac{3}{2}}e^{-\left(\lambda_{4}-\frac{\lambda_{3}^{2}}{4\lambda_{2}}\right)\left|\vec{p}\right|^{2}}\nonumber \\
 &  & \times\left(-\sqrt{\frac{3}{4\pi}}\right)\left(\frac{1}{\pi\beta_{\rho}^{2}}\right)^{\frac{3}{4}}\left(\frac{1}{\pi\beta_{\lambda}^{2}}\right)^{\frac{3}{4}},
\end{eqnarray}

\begin{eqnarray}
\Delta_{0,1} & = & \left(\frac{1}{\pi\beta_{\rho}^{'2}}\right)^{\frac{3}{4}}\left(\frac{1}{\pi\beta_{\lambda}^{'2}}\right)^{\frac{3}{4}}\left(\frac{R^{2}}{\pi}\right)^{\frac{3}{4}}\left(\frac{\pi^{2}}{\lambda_{1}\lambda_{2}}\right)^{\frac{3}{2}}e^{-\left(\lambda_{4}-\frac{\lambda_{3}^{2}}{4\lambda_{2}}\right)\left|\vec{p}\right|^{2}}\nonumber \\
 &  & \times\frac{3i}{4\pi}\left(\frac{1}{\pi\beta_{\rho}^{2}}\right)^{\frac{3}{4}}\left(\frac{8}{3\sqrt{\pi}}\right)^{\frac{1}{2}}\left(\frac{1}{\beta_{\lambda}^{2}}\right)^{\frac{5}{4}},
\end{eqnarray}

\begin{eqnarray}
    \Delta_{0,2} & = & \left(\frac{1}{\pi\beta_{\rho}^{'2}}\right)^{\frac{3}{4}}\left(\frac{1}{\pi\beta_{\lambda}^{'2}}\right)^{\frac{3}{4}}\left(\frac{R^{2}}{\pi}\right)^{\frac{3}{4}}\left(\frac{\pi^{2}}{\lambda_{1}\lambda_{2}}\right)^{\frac{3}{2}}e^{-\left(\lambda_{4}-\frac{\lambda_{3}^{2}}{4\lambda_{2}}\right)|\vec{p}|^2}\nonumber \\
 &  & \times \frac{\sqrt{15}}{8\pi}\left(\frac{1}{\pi\beta_{\rho}^{2}}\right)^{\frac{3}{4}}\left(\frac{16}{15\sqrt{\pi}}\right)^{\frac{1}{2}}\left(\frac{1}{\beta_{\lambda}^{2}}\right)^{\frac{7}{4}},
\end{eqnarray}
where the momentum space integration $\Pi(L^A_{\rho },M^A_{\rho},L^A_{\lambda},M^A_{\lambda },m)$
is introduced in Ref. \citep{Xiao2017}, what makes our differences
from Ref. \citep{Xiao2017} is that we take into account difference
of the masses of the quarks in baryons.

\end{document}